# Cascaded Channel Decoupling Based Solution for RIS Regulation Matrix

**Yajun Zhao**

**ABSTRACT** This article presents a novel solution for reconfigurable intelligent surfaces (RISs) based on cascaded channel decoupling. The proposed mechanism simplifies the RIS regulation matrix, by decomposing the electromagnetic wave regulation process into two sub-processes: virtual receiving response and virtual regular transmission, which leads to the decoupling of the RIS cascaded channel. This article further discusses the concrete implementation of the proposed channel decoupling mechanism in two scenarios of single-user access and multi-user access, and gives the corresponding detailed scheme. The numerical simulation results demonstrate that the proposed channel decoupling scheme is a low-complexity and effective solution for resolving RIS regulation matrix.

## I. INTRODUCTION

The control of electromagnetic waves has long been a goal for researchers, yet the fixed electromagnetic parameters of natural materials have limited the ability to do so to transmitters and receivers. However, the emergence of reconfigurable surfaces (RISs) has captured the interest of both academic and industrial circles due to their potential to control the wireless environment. As such, RIS technology has rapidly gained traction in both academic research and industrial applications. It is considered a promising key technology for the development of 5G-Advanced and 6G networks [1][2].

A RIS is a two-dimensional programmable metamaterial consisting of a large number of electromagnetic elements arranged periodically. By changing the state of its switching elements (such as PIN diodes, varactor diodes, and liquid crystals), it can reconstruct the arrangement of an electromagnetic response structure on the surface of an element, thereby altering the electromagnetic response characteristics [3]. The introduction of RISs has transformed the naturally uncontrollable electromagnetic propagation environment into a human-controllable one, potentially bringing about a new channel paradigm. However, accurate dynamic electromagnetic wave regulation relies on the efficient and precise solution of the RIS regulation matrix. There are many literature discussions on the channel model of RIS [4][5]. A significant body of literature offers different mechanisms for solving the RIS regulation matrix to enable channel regulation [6][7], wireless energy transmission [8][9], information modulation [10], and to address coexistence problems in RIS networks [11][12]. However, the introduction of cascaded channels after the implementation of a RIS poses significant challenges to solving the RIS regulation matrix. In our previous article [13], we briefly described the basic idea of a channel decoupling mechanism to address this challenge. In this article, we conduct an in-depth discussion and performance evaluation based on further study.

The main contributions and innovations of this article lie in the proposed cascaded channel decoupling model and algorithm for solving the RIS regulation matrix, which have the potential to significantly advance the development of RIS technology. This article proposes a new decoupling model of RIS cascaded channels, which decomposes the RIS regulation process of electromagnetic waves into two sub-processes: virtual reception response and regulation transmission. As a result, the RIS regulation matrix is decomposed into two components: the reception response matrix and the regulation transmission matrix. The decoupling model of cascaded channels achieves the decomposition of the cascaded channel into two independent segmented channels. For each segmented channel, the traditional direct path channel mechanism can be used to solve the beamforming matrix of the transmitter and receiver, thereby realizing the decoupling of RIS cascaded channels and significantly reducing the complexity of solving the RIS regulation matrix. This mechanism is referred to as the *cascaded-channel-decoupling based solution*. Based on the proposed cascaded channel decoupling model, this article discusses scenarios of single-UE access and multi-UE access, and provides corresponding solutions.

The structure of this article is as follows. Section II presents the system model and analyzes the challenge of solving the RIS regulation matrix due to the introduction of cascaded channels. Section III introduces the cascaded channel decoupling schemes for solving the RIS regulation matrix. The proposed decoupling model is further discussed in two scenarios: single-UE access and multi-UE access, with corresponding solutions provided. Section IV presents the numerical results and discussion. Finally, a conclusion is drawn.

## II. SYSTEM MODEL AND CHALLENGES

### A. SYSTEM MODEL

For the downlink of the RIS-assisted wireless communication system shown in FIGURE 1, an Node B (NB) is configured with $M$ antennas, and a RIS is configured with $N$ elements that serve users with each having $K$ antennas. Due to the introduction of the RIS into the system, the system model includes the traditional direct channel component $H_{direct} \in C^{K \times M}$ between the NB and UE and the cascaded channel $H_{cascaded} \in C^{K \times M}$ regulated by the RIS, where the cascaded channel $H_{cascaded} \in C^{K \times M}$ consists of the segmented channel $G_{nb-ris} \in C^{N \times M}$ between the NB and RIS and the segmented channel $H_{ris-ue} \in C^{K \times N}$ between the RIS and UE.

The downlink channel $H_{DL}$ can be represented as Formula (1).

$$H_{DL} = H_{cascaded} + H_{direct}$$
$$= H_{ris-ue}^{H} \Theta G_{nb-ris} + H_{direct} \quad (1)$$

where $\Theta = (\theta_1, \theta_2, \cdots, \theta_N)$ is the regulation matrix at the RIS with $\theta_n$ representing the regulation coefficient for the $n$-$th$ RIS element.

Solving blind spots is a typical application of RIS. Generally, in this case, the signal of the direct channel between NB and UE is very weak. To simplify the discussion process, it is assumed that direct channels can be ignored without losing generality (we consider further discussing the scenarios of various influencing factors, including the direct channel, in our subsequent articles). Then, Formula (1) of the RIS channel model can be rewrite as Formula (2)

$$H_{DL} = H_{cascaded}$$
$$= H_{ris-ue}^{H} \Theta G_{nb-ris} \quad (2)$$

The received signal $Y$ at the UE side can accordingly be represented as Format (3).

$$Y = H_{ris-ue}^{H} \Theta G_{nb-ris} FX + W \quad (3)$$

where $F^{M \times 1}$ is the transmitting matrix at the NB side,

$X = [x_1, x_2, ..., x_L]^T$ with satisfying $E\{XX^H\} = I$ is the transmit signal vectors of the NB, and $W \in C^{U \times 1}$ is additive white Gaussian noise at the UE side. Without losing generality, in this article, we assume that the RIS regulation is lossless and has a unitary constraint on the regulation matrix, that is, the power of the incident wave is equal to that of the reflected wave $\Theta \Theta^H = I$.

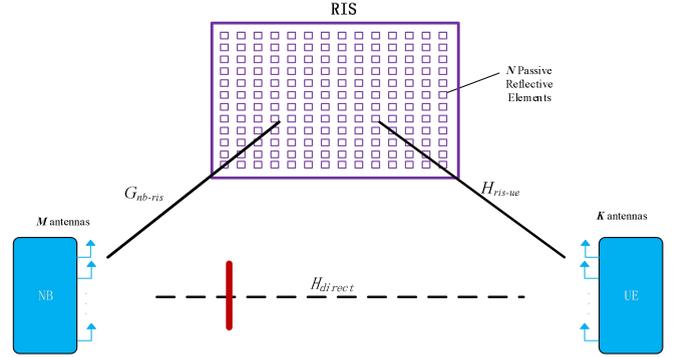

**FIGURE 1.** System model

### B. CHANLLENGES OF CASCADED CHANNELS

Formula (2) shows that the regulation coefficient matrix $\Theta$ of the RIS is located between two segmented channel matrices (i.e., the channel $G$ between the NB and RIS and the $H$ channel between the RIS and UE). It is not easy to find a modulation coefficient matrix $\Theta$ that can well match two segmented channels simultaneously by the analytical method. As seen from Formula (2), the regulation coefficient matrix of the RIS is located between two segmented channel matrices (i.e., the channel between the NB and RIS and the channel between the RIS and UE). From the point of view of mathematics, it is very difficult to directly solve the matrix components between two matrices by analytical methods for matrices with general structures. Existing literature typically employs iterative optimization or other search methods to solve for the matrix $\Theta$. In the literature [14]-[19], the beamforming matrix of the base station and the regulation matrix of the RIS are optimized by joint iteration. That is, the optimization criteria and the beam forming matrix of the base station are determined and the RIS regulation matrix is searched and optimized. Then, the RIS regulation matrix is determined, and the beamforming matrix of the base station is searched and optimized. Iterative optimization is repeated several times. However, these iterative optimization methods face many challenges, such as high computational complexity, long processing periods, and uncertain convergence rates.

In [20], the special case where a UE is equipped with a single antenna, i.e. $K = 1$, was discussed. According to the analysis of the literature, because the channel $H_{ris-ue} \in C^{1 \times N}$ between the RIS and UE is a matrix structure of $1 \times N$, in the cascaded channel model, after matrix transformation, the regulation matrix of the RIS can exchange positions with the $H_{ris-ue}$ matrix for matrix multiplication. That is, Formula (2) of the RIS channel model can be expressed as Formula (4) by matrix transformation.

$$H_{DL} = H_{ris-ue}^{H} \Theta G_{nb-ris}$$
$$= \Theta diag(H_{ris-ue}) G_{nb-ris} \qquad (4)$$

While the transformed channel model (Formula (4)) allows for obtaining the regulation matrix $\Theta$ through the decomposing matrix $HG$ using analytical means, this mathematical transformation is no longer valid when the UE is equipped with multiple antennas.

Up to now, there has been no public research on how to effectively deal with the problem of UE with multiple antennas in a simple manner, even though UEs are typically equipped with multiple antennas in practical scenarios. Therefore, it is necessary to find a simple and effective way to solve the regulation matrix $\Theta$ in such cases.

## III. CHANNEL DECOUPLING FOR SOLVING RIS REGULATION MATRIX

The above analysis shows that the existing iterative optimization solutions are highly complex and difficult to accept in practical engineering. It is necessary to find a simple method for solving the RIS regulation matrix $\Theta$ that can be applied to both single-antenna and multi-antenna scenarios.

### A. CHANNEL DECOUPLING MECHANISM

Based on the previous analysis of Formula (2) of the RIS cascaded channel model, it is evident that the RIS regulation matrix $\Theta$ is coupled between the two sub-channel matrices, resulting in high complexity when solving for the RIS regulation matrix $\Theta$. A natural idea is that if the cascaded channel can be decoupled and the two sub-channels of the cascaded channel can be decomposed separately, which can significantly reduce the complexity of the solution.

Taking a new perspective on the RIS system model, it is evident that the regulation of incident electromagnetic waves by RIS can be further divided into two virtual sub-processes: the receiving response sub-process and the output regulation sub-process. However, existing solution mechanisms treat the regulation of RIS as a single process, resulting in the need to solve the coupled segmented channels of the cascaded channels together.

Analyzing the above two virtual sub-processes, it becomes clear that the receiving response virtual sub-process of RIS can be viewed as a virtual receiver. The optimal design of this virtual receiver can be based on the principles of traditional optimal receiver design, depending on the channels that the incident signal has experienced. Since RIS is a multi-antenna array, the optimization of this virtual receiver can be transformed into the optimization of the optimal receiving matrix. After the incident signal is processed by the virtual response sub-process, it undergoes channel equalization processing and is transformed into the original signal without channel distortion, before being mapped to the outgoing antennas one by one. The output regulation sub-process of RIS can be regarded as a virtual transmitter with a virtual multi-antenna array. The optimal design of this virtual transmitter can be optimized based on the principles of traditional optimal transmitter design and depends on the channel between this virtual transmitter and its corresponding receiver. As this virtual transmitter is a multi-antenna array, the optimal transmitter design is transformed into the optimal transmit precoding matrix design for the RIS antenna array.

Based on the above analysis, the optimal processing matrices for the two virtual sub-processes of RIS regulation can be expressed as the receiving response matrix $\Theta_2$ of RIS and the output regulation matrix $\Theta_1$ of RIS, respectively. The receiving response matrix $\Theta_1$ corresponds to the sub-channel $G_{nb-ris}$ between the transmitter (NB) and RIS, while the output regulation matrix $\Theta_2$ corresponds to the segmented channel $H_{ris-ue}^{H}$ between RIS and the receiver (UE). In other words, the receiving response matrix $\Theta_1$ can be calculated based on the segmented channel $G_{nb-ris}$, and the output regulation matrix $\Theta_2$ can be calculated based on the segmented channel $H_{ris-ue}^{H}$. Based on this idea, the complete regulation matrix $\Theta$ of RIS can be decomposed into two sub-matrix components: the receiving response sub-matrix $\Theta_1$ and the reflection regulation sub-matrix $\Theta_2$.

Taking the downlink (DL) channel as an example, the series formulas for the sub-process of the cascaded channel correlation processing of RIS are as follows.

(a) Expression for RIS Receiving Response Virtual Sub-process

The downlink signal $Y_{DL,ris,in}$ incident on RIS through the channel between NB and RIS is denoted as (5) before being processed by RIS.

$$Y_{DL,ris,in} = GFX \qquad (5)$$

The signal X expression (6) of the downlink signal Y incident on the RIS panel is processed by the RIS virtual response sub-process (also known as the virtual receiving sub-process).

$$X_{ris} = \Theta_1 Y_{DL,ris,in} = \Theta_1 GFX \qquad (6)$$

In Formulas (5) and (6), $F$ is the multi-antenna precoding matrix of NB side; $X = [x_1, x_2, ..., x_L]^T$ is the data sequence sent by NB, satisfying $E\{XX^H\} = I$. Since RIS is passive regulation, there is no Additive white Gaussian noise term in Formula (5).

(b) Expression for RIS Output Modulation Virtual Sub-process

The signal $X_{ris}$ processed by the RIS virtual response sub-process is further processed by the RIS virtual regulation sub-process (also called the virtual transmission sub-process), and the processed output signal $X_{DL,ris,out}$ is expressed as (7a).

$$X_{DL,ris,out} = \Theta_2 X_{ris} \qquad (7a)$$

The formula (6) is introduced into Formula (7a), and Formula (7b) is obtained.

$$X_{DL,ris,out} = \Theta_2 \Theta_1 G_{nb-ris} FX \qquad (7b)$$

The signal $X_{DL,ris,out}$ processed by the two virtual subprocesses of RIS flows to the UE channel through the RIS, so the downlink signal $Y_{DL,ue}$ at the UE receiving end is expressed as Formula (8)

$$Y_{DL,ue} = H_{ris-ue}^H X_{DL,ris,out} + W \qquad (8)$$

Where, $H_{ris-ue}^H$ is the channel between RIS and UE; $W$ is the additive Gaussian white noise matrix of UE.

(c) To combine the above sub-processes, and get a complete formula

By combining the above Formulas (5) to (8), Formula (9) can be obtained. Formula (9) is the expression for the whole process of the downlink signal reaching the UE receiver after being processed by two virtual sub processes of RIS.

$$Y_{DL,ue} = H_{ris-ue}^H \Theta_2 \Theta_1 G_{nb-ris} FX + W \qquad (9)$$

Comparing Formulas (3) and (9), the expression (10) of RIS regulation matrix can be obtained naturally.

$$\Theta = \Theta_2 \Theta_1 \qquad (10)$$

Then Formula (2) of the downlink cascaded channel model can be written as Formula (11).

$$H_{DL} = H_{ris-ue}^H \Theta_2 \Theta_1 G_{nb-ris} \qquad (11)$$

The above idea is similar to the analog beamforming matrix of the transmitter and the analog beamforming response matrix of the receiver in traditional large-scale MIMO hybrid beamforming [21]. However, in traditional large-scale MIMO scenarios, this type of processing involves a direct decomposition of the direct path channel between NB and UE. In the cascaded channel model of RIS, the regulation matrix $\Theta$ of RIS cannot be obtained by directly decomposing the channel matrix. A new perspective is needed, where the regulatory behavior of RIS is assumed to be handled as the two virtual sub-processes mentioned above.

**Remark 1:** The regulation process of RIS is virtualized into two sub-processes: the receiving response sub-process (or virtual receiving sub-process) for the incident electromagnetic waves and the output regulation sub-process (or virtual transmitting sub-process) for the emitted electromagnetic wave. The two virtual sub-processes can correspond to two segmented channels of the cascaded channel model, respectively. Then, two virtual processing sub-matrices can be obtained by independently decomposing two segmented channel matrices. This method achieves the decoupling of RIS cascaded channels and greatly reduces the complexity of solving the RIS regulation matrix.

**Remark 2:** It should be noted that Formula (5) needs to meet at least two conditions: firstly, the two segmented channels $H_{ris-ue}$ and $G_{nb-ris}$ can be decomposed into a receiving response matrix and a transmitting regulation matrix, respectively. Additionally, due to the passive regulation of RIS and the assumption of no energy loss during the regulation process, the regulation matrix usually needs to satisfy constant modulus constraints.

Next, we will further explore specific solution examples of $\Theta_1$ and $\Theta_2$ for single-user access scenarios and multi-user access scenarios, respectively.

### B. CHANNEL DECOUPLING FOR SINGLE-USER ACCESS

The above formulas (5) - (11) can be directly used for single-user access scenarios. Here, we will first discuss the application of the proposed scheme in a single-user access case.

To facilitate discussion, without losing generality, singular value decomposition (SVD) is used to analyze the solution process of $\Theta_1$ and $\Theta_2$, although other algorithms for solving $\Theta_1$ and $\Theta_2$ are also applicable. As mentioned above, the basic solution process is to decompose the two segmented channels of the RIS cascaded channel by SVD, independently solve the two virtual sub-process regulation matrices $\Theta_1$ and $\Theta_2$ of RIS, and then obtain the complete regulation matrix of RIS according to Formula (10).

The SVD decomposition of two segmented channels $G_{nb-ris}$ and $H_{ris-ue}^H$ can be expressed as Formula (12).

$$SVD(G_{nb-ris}) = U_G D_G V_G^H \qquad (12a)$$

$$SVD(H_{ris-ue}^H) = U_H D_H V_H^H \qquad (12b)$$

The receiving matrix of segmented channel $G_{nb-ris}$ is $V_G$, and that of segmented channel $H_{ris-ue}^H$ is $U_H^H$.

Compared with Formula (10), if $\Theta_1 = U_G^H$ and $\Theta_2 = V_H$ are given, Formula (10) can be transformed into Formula (13).

$$\Theta = \Theta_2 \Theta_1 = V_H U_G^H \qquad (13)$$

If Formula (13) is introduced into Formula (11), it will be updated and represented as Formula (14).

$$\begin{aligned} H_{DL} &= H_{ris-ue}^H \Theta_2 \Theta_1 G_{nb-ris} \\ &= H_{ris-ue}^H V_H U_G^H G_{nb-ris} \end{aligned} \qquad (14)$$

By combining Formula (14) with Formula (9), the received signal $Y_{DL,ue}$ at the UE side can be expressed as Formula (15).

$$Y_{DL,ue} = H_{ris-ue}^H V_H U_G^H G_{nb-ris} FX + W \qquad (15)$$

In addition, SVD is a unitary transformation operation of the matrix, which meets the constant modulus criterion of the RIS regulation matrix. That is, if $\Theta_1 = U_G^H$ is a unitary matrix and $\Theta_2 = V_H$ is a unitary matrix, then $\Theta = V_H U_G^H$ is also a unitary matrix.

The achievable rate $R$ of a RIS-assisted MIMO system is given in Formula (16).

$$R = B \log_2 det(I_K + \frac{\rho}{N\sigma^2} H_{DL} F F^H H_{DL}^H) \quad (16)$$

According to the previous discussion, the relevant variable expressions of Formula (10) are substituted. Then, Formula (14) can be expressed as Formula (15).

$$R = B \log_2 det(I_K + \frac{\rho}{N\sigma^2}(H_{ris-ue} V_H U_G^H G_{nb-ris}) F F^H (H_{ris-ue} V_H U_G^H G_{nb-ris})^H) \quad (17)$$

where $B$ is the occupied frequency bandwidth, $\rho$ is the transmission power of the NB, and $\sigma$ is the variance of Gaussian white noise corresponding to $W$.

Another approach is to use traditional classical optimization methods to solve the regulation matrix components $\Theta_1$ and $\Theta_2$ of the two virtual regulation sub-processes, respectively. There has been a lot of discussion about related optimization algorithms in existing literature. This article only provides the optimization problem and its constraints, as shown in Formula (18). Details of specific solutions will not be repeated here.

$$P1: \max_{\forall \vartheta_i} R_{sum} = \sum_{k=1}^{K} f(G_k, H_k, F_k, p_k)$$
$$St. \ \|\theta_i\| = 1, \ \forall \theta_i$$
$$\sum_{k}^{K} p_k \le p$$
$$\|F_k\| = 1, \ k = 1,...K \quad (18)$$

Where, $G_k$, $H_k^H$ and $F_k$ are the channel components of $u_k$ and the multi-antenna precoding matrix on the NB side for the $u_k$ 's signal, respectively; $\theta_i$ is the regulation coefficient of the $ith$ element of RIS; $p_k$ is the allocated power for $u_k$, satisfying $\sum_{k=1}^{K} p_k \le p_{max}$; $K$ is the total number of UE, and $p_{max}$ is the maximum rated transmission power of the base station.

Get $\Theta_1 = \Phi_{1,opt}$ and $\Theta_2 = \Phi_{2,opt}$ respectively, and then get expression (19) according to formula (10).

$$\Theta = \Theta_2 \Theta_1 = \Phi_{2,opt} \Phi_{1,opt} \quad (19)$$

## C. CHANNEL DECOUPLING FOR MULTI-USER ACCESS

For multi-user access scenarios, without loss of generality, this article will discuss multi-user access using OFDMA and take two UEs, $ue_1$ and $ue_2$, as examples. Due to the orthogonality of the frequency domain, it can be assumed that the channels of the two UEs are independent, i.e., $G_{ue_1}$ and $G_{ue_2}$ are independent, and $H_{ue_1}$ and $H_{ue_2}$ are independent. The PA-based multi-beam mechanism described in reference [22] can be used to solve the multi-user RIS regulation matrix. That is, the regulation matrices $\Phi_{ue_1}$ and $\Phi_{ue_2}$ of $ue_1$ and $ue_2$ are solved respectively, and then the complete regulation matrix of RIS for two UEs is obtained using the PA mechanism. Without loss of generality, SVD is still used to solve the regulation matrices of $ue_1$ and $ue_2$.

The expressions to solve the regulation matrix corresponding to $ue_1$ are given by Formulas (20 a/b) and (21).

$$SVD(G_{ue1}) = U_{G,ue1} D_{G,ue1} V_{G,ue1}^H \quad (20a)$$
$$SVD(H_{ue_1}^H) = U_{H,ue1} D_{H,ue1} V_{H,ue1}^H \quad (20b)$$
$$\Theta_{ue_1} = \Theta_{ue_1,2} \Theta_{ue_1,1} = V_{H,ue_1} U_{G,ue_1}^H \quad (21)$$

The expressions for solving the corresponding regulation matrix of $ue_2$ are formulas (22 a/b) and (23).

$$SVD(G_{ue2}) = U_{G,ue2} D_{G,ue2} V_{G,ue2}^H \quad (22a)$$
$$SVD(H_{ue_2}^H) = U_{H,ue2} D_{H,ue2} V_{H,ue2}^H \quad (22b)$$
$$\Theta_{ue_2} = \Theta_{ue_2,2} \Theta_{ue_2,1} = V_{H,ue_2} U_{G,ue_2}^H \quad (23)$$

The formula (4.15) based on the PA mechanism is used for two UE, $ue_1$ and $ue_2$. The complete regulation matrix $\Theta$ of RIS is formula (24).

$$\Theta_{mu} = \sum_{k=1}^{K} \alpha_k \Theta_k \quad (24)$$

Where, $\Theta_k$ represents the RIS regulation matrix corresponding to the single connection of $ue_k$; The complex number $\alpha_k = \rho_k \exp(-j\vartheta_k)$ represents the weighting coefficient of the regulation matrix component of $ue_k$ satisfying constraint $\sum_{k=1}^{K} \|\alpha_k\|_2 \le 1$; $K$ represents the total number of UE connected simultaneously.

For $K=2$, the formulas (21) and (23) are introduced into formula (24), and the final expression of the RIS regulation matrix of $ue_1$ and $ue_2$ is obtained formula (25).

$$\Theta_{mu} = \alpha_{ue_1} \Theta_{ue_1} + \alpha_{ue_2} \Theta_{ue_2} \quad (25)$$

## IV. NUMERICAL SIMULATION

In the preceding sections, we conducted a comprehensive theoretical analysis of the channel decoupling scheme. Building on these analyses, this section employs the Monte Carlo method to numerically evaluate the proposed method in both single-user and multi-user access scenarios.

### A. PERFORMANCE EVALUATION OF SINGLE-USER CASES

In this simulation, we assume a far-field channel model and employ a uniform planar array (UPA) with $M = M_x \times M_y$ elements for the RIS, while fixing $M_x = 50$ and increasing $M_y \in [1,2,4,8,16] \times 2$. The base station (BS) uses a Uniform Planar Array (UPA) with $N = N_x \times N_y$ elements, $N_x = 8$ and $N_y = 4$. The carrier frequency is set to $f_c = 28GHz$.

We compare the proposed cascaded channel decoupling algorithm (refers to Formula 17) with the traditional typical alternating optimization (AO) algorithm and take the random phase regulation method as the baseline. This random phase models the random scattering effect on the surface of natural scatters when the RIS is not deployed.

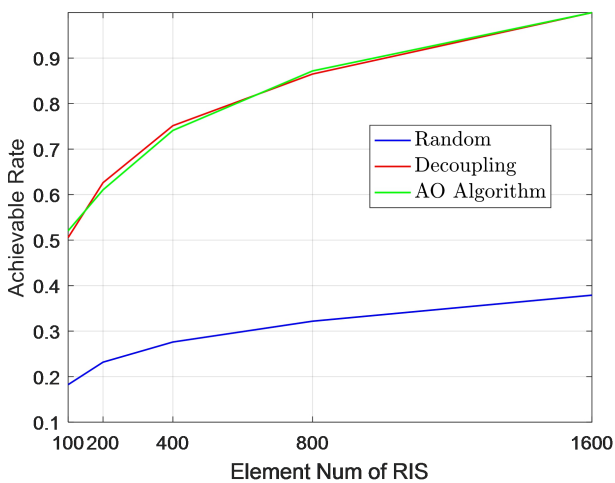

**FIGURE 2.** Performance of single-user cases (note, this figure shows the normalized data rates.)

FIGURE 2 depicts the achievable data rate versus the number $M$ of RIS elements, where the transmission power $p$ is fixed and $M$ varies. The "Random" curve indicates the performance achieved when the incident electromagnetic wave is randomly scattered; the "Decoupling" curve indicates the performance achieved when the proposed decoupling schemes are used. From the simulation curves, it can be seen that the proposed algorithm can achieve the equivalent performance as the traditional AO algorithm, and both have significant RIS beamforming gain relative to the random phase. The simulation results show that the proposed RIS cascaded channel decoupling mechanism can be used as an effective method to solve the RIS regulation matrix with low complexity.

### B. PERFORMANCE EVALUATION OF MULTI-USER CASES

Assuming two UEs are utilizing OFDMA multiple access and PA mechanism for multi-beam, while using a far-field model channel. The base station's antenna type is ULA, with $M = 64$ antennas, and the RIS antenna type is ULA with $N \in \{800, 1600, 3200, 6400\}$ antenna elements.

Additionally, the central frequency is $f_c = 5GHz$, and the number of users is $K = 2$.

With evaluating the PA-based multi-UE access performance using formula (23), three performance cases are provided for comparison - perfect regulation, unexpected exception regulation, and simple signal mirroring. As shown in the numerical simulation curve in FIGURE 3, while the PA mechanism does have a slight performance loss compared to the ideal situation, it still provides better performance gains compared to unexpected regulation and simple mirroring.

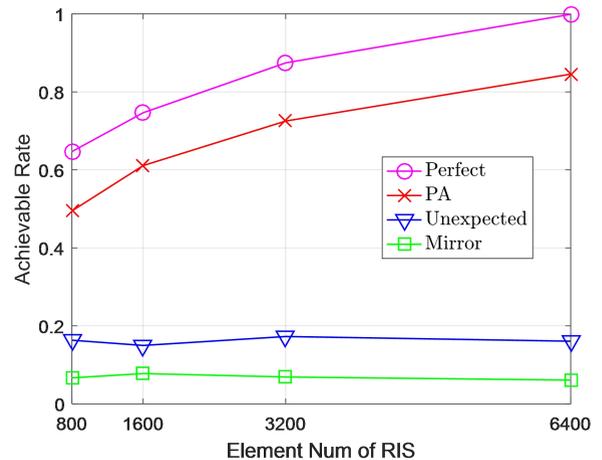

**FIGURE 3.** Performance of multi-user cases ("perfect" means the best performance without considering the constraints specified by RIS and represents the upper bound of performance; "Passive RIS PA" indicates the use of PA mechanism; "Unexpected" indicates that the signal is subject to unexpected abnormal regulation; and "Mirror" indicates simple mirror regulation of the signal.)

The above numerical simulation results show that the proposed RIS cascaded channel decoupling mechanism is an effective method to solve the RIS regulation matrix with low complexity.

### V. CLONCLUSION

This article proposes a new decoupling model for the RIS cascaded channel, which decomposes the electromagnetic wave regulation process by the RIS into two sub-processes: virtual receiving response and virtual regular transmission. This decomposition divides the RIS regulation matrix into two components: the receiving response matrix and the regulation transmission matrix. By using this cascaded channel decoupling mechanism, two segmented channels can be solved independently, greatly simplifying the calculation of the RIS regulation matrix. The article discusses the proposed decoupling mechanism in two scenarios: single UE access and multi-UE access, providing corresponding solutions. The numerical simulation results demonstrate that the proposed channel decoupling scheme is a low-complexity and effective solution for solving the RIS regulation matrix.

Future research is needed on the specific implementation and performance evaluation of different sub-control matrix methods for solving segmented channels. Additionally, further research is also needed on the specific implementation and performance evaluation of RIS channel models that store cascading and direct channel components simultaneously.